\documentstyle[12pt,aps,floats]{revtex}
\begin{document}
\topmargin -2cm
\textwidth 20cm
\author{Mauricio Cataldo$^1\thanks{%
mcataldo@zeus.dci.ubiobio.cl}$ and Patricio Salgado$^2\thanks{%
Present adress: Sektion Physik, L.M.Universt\"at M\"unchen. Theresienstra$%
\beta $e 37, D-80333, M\"unchen, Germany.}\thanks{%
psalgado@halcon.dpi.udec.cl}$}
\address{$^1$Departamento de F\'\i sica, Facultad de Ciencias,
Universidad del B\'{\i}o-B\'{\i}o, Casilla 5-C, Concepci\'{o}n, Chile. \\
$^2$Departamento de F\'\i sica, Facultad de Ciencias F\'\i s.y Mat.,
Universidad de Concepci\'on, Casilla 4009, Concepci\'on, Chile.\\
\smallskip\ }
\title{The Einstein-Maxwell Extreme BTZ Black Hole with Self (Anti-self) Dual
Maxwell Field. }
\maketitle
\begin{abstract}
Recently M. Kamata and T. Koikawa claimed to obtain the
charged version of the spinning BTZ black hole solution by assuming an
(anti-) self dual condition imposed on the orthonormal basis components of
the electric and magnetic fields. We point out that the Kamata-Koikawa field
is not a solution of the Einstein-Maxwell equations and we find the correct
solution of the studied problem. A spinning magnetic solution is also found.
It is shown that a duality mapping exists among spinning solutions obtained
from electrostatic and magnetostatic fields with the help of the local
transformation $t\longrightarrow t-\omega \,\theta ,\,\,\,\theta
\longrightarrow \theta -\omega t.$\\ %}


\end{abstract}
\smallskip\ 
It is well known that 2+1 dimensional general relativity has counterpart to
the four dimensional Kerr black hole. This three dimensional spinning black
hole was found by Ba\~nados, Teitelboim and Zanelli (BTZ)~\cite
{Teitelboim1,Teitelboim2}. The BTZ solution, with mass $M$ and angular
momentum $J$, may be written in the form 
\begin{eqnarray}
ds^2=N^2dt^2-L^{-2}dr^2-K^2(N^\phi dt+d\phi )^2,  \label{metrica Kamata}
\end{eqnarray}
with the following metric functions: 
\begin{eqnarray}
\label{BTZ}
N^2 =L^2=-M-\Lambda r^2+J^2/4r^2,\,\,\,K^2=r^2,\,\,\, N^\phi =A\pm J/2r^2,
\end{eqnarray}
where $\Lambda <0$ is the cosmological constant and $A$ is an arbitrary
constant. The first attempt to obtain the 2+1 Kerr-Newman counterpart was
made by the same authors~\cite{Teitelboim1,Teitelboim2}. The mentioned field is
a solution only for the Einstein equations but not for the Maxwell equations
without sources.

Recently, Kamata and Koikawa~\cite{Kamata} made another attempt and found an
Einstein-Maxwell field, with a negative cosmological constant $\Lambda $.
This solution was obtained by assuming a self (anti-self) dual equation $E_{%
\hat r}=\pm B_{\,\,\hat {}}$, which is imposed on the orthonormal basis
components of the electric field $E_{\hat r}$ and the magnetic field $B_{\,\,%
\hat {}}$. The Einstein-Maxwell action considered in their paper is 
\begin{eqnarray}
S=\frac 1{16\pi G}\int \sqrt{-g}(R-2\Lambda -4\pi GF^2)\,d^3x,
\end{eqnarray}
where $G$ is Newton's constant, $\Lambda $ is the cosmological constant and $%
F^2\equiv g^{\mu \nu }g^{\rho \sigma }F_{\mu \rho }F_{\nu \sigma }$.

The solution proposed in~\cite{Kamata} can be written in the form~(\ref
{metrica Kamata}), where now 
\begin{eqnarray}  \label{Kamata-Koikawa}
N^2 &=&\frac{r^2}{K^2}L^2,\,\,\,N^\phi =-\frac J{2K^2}\left( 1+\frac{K^2-r^2%
}{r_0^2}\right) ,  \nonumber \\
L^2 &=&-8\pi GQ^2+\Lambda r^2+\frac{J^2}{4r^2},\hspace{1.8cm} \\
K^2 &=&r^2+r_0^2\,ln\left| \frac{r^2-r_0^2}{r_0^2}\right| .\hspace{2cm} 
\nonumber
\end{eqnarray}
$Q$ is the magnitude of the electric charge, $r_0^2=4\pi GQ^2/\Lambda $ and $%
J=\varepsilon 8\pi GQ^2/\sqrt{\Lambda }$. %We see that when $Q=0$

Unfortunately, the Einstein-Maxwell equations obtained by the authors are
not completely correct . In fact, a careful study of the self-consistent
equations shows that equation (14) of~\cite{Kamata} 
\begin{eqnarray}
R_{\hat t\hat \phi }=L(\beta ^{^{\prime }}+2K^{-1}K^{^{\prime }}\beta
)=-8\pi GE_{\hat r}B_{\,\,\hat {}},  \nonumber
\end{eqnarray}
is not correct, ($\beta =-KLN^{-1}N^{\phi ^{^{\prime }}}/2$). The minus sign
on the right hand side of this equation must be plus. Besides meticulous
derivation of the equations, one can easily check this sign by computing our
solution~(\ref{cat}) with $M=\Lambda =0$.

In this Letter we obtain the correct charged version of the
(2+1)-dimensional spinning BTZ black hole by assuming an (anti-) self dual
condition imposed on the electromagnetic field. Firstly, we obtain the
Einstein-Maxwell equations for the general rotating case following the
procedure used in~\cite{Cataldo}, and afterwards we deduce the equations
considering the (anti-) self dual condition.

We write the line element of the Einstein-Maxwell field in the form 
\begin{eqnarray}
ds^2=e^{2\alpha (r)}dt^2-e^{2\beta (r)}dr^2-(e^{\gamma (r)}d\theta
+e^{\delta (r)}dt)^2.  \label{metrica}
\end{eqnarray}
Here we adopt the following four potential of the electromagnetic field: 
\begin{eqnarray}
A=A_a(r)\,dx^a=q_0A_0(r)\,dt+q_2A_2(r)\,d\theta ,  \label{metr}
\end{eqnarray}
where $A_0$ and $A_2$ are arbitrary functions of the r coordinate, and the
constant coefficients $q_0$ and $q_2$ are introduced for switching off the
electric and/or magnetic fields.

From $F=dA$ and (\ref{metr}) we have 
\begin{eqnarray}
F=dA=E_r\,dr\,\wedge \,dt+B\,dr\,\wedge \,d\theta ,  \label{metri}
\end{eqnarray}
where $E_r=q_0\,A_0^{\prime }$ and $B=q_2\,A_2^{\prime }$ and the
differentiation with respect to r is denoted by ${}^{\prime }$. To write the
self-consistent equations we introduce an orthonormal 1-form basis 
\begin{eqnarray}
\omega ^{\hat t}=e^\alpha \,dt,\,\,\,\omega ^{\hat r}=e^\beta
\,dr,\,\,\,\omega ^{\hat \theta }=e^\gamma \,d\theta +e^\delta \,dt.
\label{tetrada}
\end{eqnarray}
In this basis the non-trivial components of the Einstein tensor are (the
definition $R_{a b}=R^{c}{}_{abc}$ is used here)
\begin{eqnarray}
G_{\hat t\hat t}=e^{-2\beta }\left( \gamma {}^{\prime \prime }+\gamma
{}^{\prime }{}^2-\beta {}^{\prime }\gamma {}^{\prime }\right) +\frac 14I,
\end{eqnarray}
\begin{eqnarray}
G_{\hat r\hat r}=-\alpha {}^{\prime }\gamma {}^{\prime }e^{-2\beta }-\frac 14%
I,
\end{eqnarray}
\begin{eqnarray}
G_{\hat \theta \hat \theta }=e^{-2\beta }\left( \alpha {}^{\prime }\beta
{}^{\prime }-\alpha {}^{\prime \prime }-\alpha {}^{\prime }{}^2\right) +%
\frac 34I,
\end{eqnarray}
and 
\begin{eqnarray}
G_{\hat t\hat \theta } =\frac 12e^{-\alpha -2\beta +\delta }
\left( \gamma {}^{\prime \prime }-\delta {}^{\prime \prime }+(\alpha
{}^{\prime }+\beta {}^{\prime })(\delta {}^{\prime }-\gamma {}^{\prime
})-\delta {}^{\prime }{}^2+2\gamma {}^{\prime }{}^2-\gamma {}^{\prime
}\delta {}^{\prime }\right) ,
\end{eqnarray}
where
\begin{eqnarray}
I=(\gamma {}^{\prime }-\delta {}^{\prime })^2e^{-2\alpha -2\beta +2\delta }.
\end{eqnarray}
The energy-momentum tensor of the electromagnetic field, in Gaussian units,
is given by 
\begin{eqnarray}
T_{\hat a\hat b}=\frac{g_{\hat a\hat b}}{8\pi }\,F_{\hat c\hat d}\,F^{\hat c%
\hat d}-\frac 1{2\pi }F_{\hat a\hat c}\,F_{\hat b}^{\,\,\,\hat c}.
\label{energia-momentum}
\end{eqnarray}
To gets its components we must compute $F_{\hat a\hat b}$. In the
orthonormal basis~(\ref{tetrada}) we have 
\begin{eqnarray}
F=E_{\hat r}\,\omega ^{\hat r}\,\wedge \,\omega ^{\hat t}+B_{\,\,\hat {}%
}\,\omega ^{\hat r}\,\wedge \,\omega ^{\hat \theta },
\end{eqnarray}
where 
\begin{eqnarray}
E_{\hat r}=E_re^{-\alpha -\beta }-Be^{-\alpha -\beta -\gamma +\delta
},\,\,\,B_{\,\,\hat {}}=Be^{-\beta -\gamma }.
\end{eqnarray}
Thus from~(\ref{energia-momentum}) we get 
\begin{eqnarray}
T_{\hat a\hat b}^{e.m.}\,\omega ^{\hat a}\,\otimes \,\omega ^{\hat b}
&=&\left( \frac 1{4\pi }\,E_{\hat r}{}^2+\frac 1{4\pi }\,B_{\,\,\hat {}%
}{}^2\right) \,\omega ^{\hat t}\,\otimes \,\omega ^{\hat t}+  \nonumber \\
&&\left( -\frac 1{4\pi }\,E_{\hat r}{}^2+\frac 1{4\pi }\,B_{\,\,\hat {}%
}{}^2\right) \,\omega ^{\hat r}\,\otimes \,\omega ^{\hat r}+  \nonumber \\
&&\left( \frac 1{4\pi }\,E_{\hat r}{}^2+\frac 1{4\pi }\,B_{\,\,\hat {}%
}{}^2\right) \,\omega ^{\hat \theta }\,\otimes \,\omega ^{\hat \theta }+ 
\nonumber \\
&&\frac 1{2\pi }E_{\hat r}\,B_{\,\,\hat {}}\left( \omega ^{\hat t}\,\otimes
\,\omega ^{\hat \theta }+\omega ^{\hat \theta }\,\otimes \,\omega ^{\hat t%
}\right) .\hspace{1cm}  \nonumber
\end{eqnarray}
It is clear that the source-free Maxwell's equations $(\sqrt{-g}%
F^{ab})_{,b}=0$ are satisfied if 
\begin{eqnarray}
E_r\,e^{-\alpha -\beta +\gamma }-B\,e^{-\alpha -\beta +\delta }=\tilde {C_0},
\label{cond de Max1}
\end{eqnarray}
and 
\begin{eqnarray}
E_re^{-\alpha -\beta +\delta }+B\,e^{-\alpha -\beta -\gamma }(e^{2\alpha
}-e^{2\delta })=\tilde {C_2},  \label{cond de Max2}
\end{eqnarray}
where $\tilde C_0$ and $\tilde C_2$ are constants of integration. The
Einstein equations are given by 
\begin{eqnarray}
G_{\hat a\hat b}+\Lambda g_{\hat a\hat b}=-\kappa T_{\hat a\hat b}.
\end{eqnarray}
Thus the self-consistent equations are formed by 
\begin{eqnarray}
G_{\hat t\hat t}=-\frac \kappa {4\pi }\left( E_{\hat r}{}^2+B_{\,\,\hat {}%
}{}^2\right) -\Lambda ,  \label{t-t1}
\end{eqnarray}
\begin{eqnarray}
G_{\hat r\hat r}=\frac \kappa {4\pi }\left( E_{\hat r}{}^2-B_{\,\,\hat {}%
}{}^2\right) +\Lambda ,
\end{eqnarray}
\begin{eqnarray}
G_{\hat \theta \hat \theta }=-\frac \kappa {4\pi }\left( E_{\hat r%
}{}^2+B_{\,\,\hat {}}{}^2\right) +\Lambda ,
\end{eqnarray}
\begin{eqnarray}
G_{\hat t\hat \theta }=-\frac \kappa {2\pi }E_{\hat r}B_{\,\,\hat {}},
\label{t-theta1}
\end{eqnarray}
together with the Maxwell conditions~(\ref{cond de Max1})--(\ref
{cond
de
Max2}).

Now we assume the self dual (anti-self dual) equation on the orthonormal
basis components of the electric and the magnetic fields 
\begin{eqnarray}
E_{\hat r}=\varepsilon B_{\,\,\hat {}},\,\,\varepsilon =\pm 1.
\label{dual-antidual}
\end{eqnarray}
Thus the Einstein equations take the form 
\begin{eqnarray}
G_{\hat t\hat t}=-\frac \kappa {2\pi }B_{\,\,\hat {}}{}^2-\Lambda ,
\label{Gtt}
\end{eqnarray}
\begin{eqnarray}
G_{\hat r\hat r}=\Lambda ,  \label{Grr}
\end{eqnarray}
\begin{eqnarray}
G_{\hat \theta \hat \theta }=-\frac \kappa {2\pi }B_{\,\,\hat {}%
}{}^2+\Lambda ,  \label{Gthetatheta}
\end{eqnarray}
\begin{eqnarray}
G_{\hat t\hat \theta }=-\frac \kappa {2\pi }\varepsilon B_{\,\,\hat {}}{}^2,
\label{Gttheta}
\end{eqnarray}
and the Maxwell conditions take respectively the form 
\begin{eqnarray}
Be^{-\beta }=C_0,  \label{Maxwell condition 1}
\end{eqnarray}
\begin{eqnarray}
Be^{-\beta -\gamma }\left( \varepsilon \,e^\delta +e^\alpha \right) =C_2,
\label{Maxwell condition 2}
\end{eqnarray}
where $C_0=\varepsilon \,\tilde {C_0}$ and $C_2=\tilde {C_2}$. From~(\ref
{Maxwell condition 1})--(\ref{Maxwell condition 2}) we have 
\begin{eqnarray}
C_0\left( \varepsilon \,e^\delta +e^\alpha \right) =C_2\,e^\gamma .
\label{Maxwell condition 3}
\end{eqnarray}
By derivating the expression~(\ref{Maxwell condition 3}) we obtain 
\begin{eqnarray}
\varepsilon \,(\delta ^{^{\prime }}-\gamma ^{^{\prime }})e^\delta =(\gamma
^{^{\prime }}-\alpha ^{^{\prime }})e^\alpha  \label{condicion derivada}
\end{eqnarray}
which implies that 
\begin{eqnarray}
I=(\gamma ^{^{\prime }}-\alpha ^{^{\prime }})^2\,e^{-2\beta }.  \label{I}
\end{eqnarray}
From~(\ref{Grr}) and (\ref{I}) we obtain 
\begin{eqnarray}
\alpha ^{^{\prime }}+\gamma ^{^{\prime }}=2\,\sqrt{-\Lambda }\,e^\beta ,
\label{lambda negativo}
\end{eqnarray}
which implies that the solution is valid only for $\Lambda \leq 0$.
Subtracting equations~(\ref{Gtt}), (\ref{Gthetatheta}) and using~(\ref{I})
and (\ref{lambda negativo}) we get the identity $0\equiv 0$. Adding
equations~(\ref{Gtt}), (\ref{Gthetatheta}) and using~(\ref{I}) and (\ref
{lambda
negativo}) we get the second-order differential equation for $%
e^\gamma $ 
\begin{eqnarray}
2\gamma ^{^{\prime \prime }}+4\gamma ^{^{\prime }}{}^2-2\beta ^{^{\prime
}}\gamma ^{^{\prime }}-4\sqrt{-\Lambda }\gamma ^{^{\prime }}e^\beta =-\frac 
\kappa \pi Be^{-2\gamma }.
\end{eqnarray}
If we take, for $g_{rr}$, the gauge $dr^2$ instead of $e^{2\beta (r)}dr^2$
(i.e $e^{2\beta }=1$), we have the following equation: 
\begin{eqnarray}
(e^{2\gamma })^{^{\prime \prime }}-2\sqrt{-\Lambda }(e^{2\gamma })^{^{\prime
}}=-\frac \kappa \pi C_0{}^2,
\end{eqnarray}
and hence 
\begin{eqnarray}
e^{2\gamma }=\frac{\kappa \,C_0{}^2}{2\pi \sqrt{-\Lambda }}\,r+Ce^{2\sqrt{%
-\Lambda }\,r}+D,
\end{eqnarray}
where $C$ and $D$ are constants of integration. Now adding (\ref{Gtt})+(\ref
{Gthetatheta})+(\ref{Gttheta}) we obtain 
\begin{eqnarray}
(e^{\gamma +\delta })^{^{\prime \prime }}-2\sqrt{-\Lambda }(e^{\gamma
+\delta })^{^{\prime }}=-\frac{\kappa \,C_0\,C_2\,\varepsilon }\pi ,
\end{eqnarray}
and thus 
\begin{eqnarray}
e^{\gamma +\delta }=\frac{\kappa C_0C_2\varepsilon }{2\pi \sqrt{-\Lambda }}%
\,r+Ee^{2\sqrt{-\Lambda }\,r}+F.
\end{eqnarray}
This means that 
\begin{eqnarray}
e^\delta =\frac{\frac{\kappa C_0C_2\varepsilon }{2\pi \sqrt{-\Lambda }}%
\,r+Ee^{2\sqrt{-\Lambda }\,r}+F}{\sqrt{\frac{\kappa \,C_0{}^2}{2\pi \sqrt{%
-\Lambda }}\,r+Ce^{2\sqrt{-\Lambda }\,r}+D}}.
\end{eqnarray}
Integrating~(\ref{lambda negativo}) we find that $e^\alpha $ is given by 
\begin{eqnarray}
e^\alpha =Ge^{2\sqrt{-\Lambda }\,r}\,e^{-\gamma }.
\end{eqnarray}
Lastly, from~(\ref{Maxwell condition 3}) we obtain the following relation
for the constants of integration: 
\begin{eqnarray}
C_0\left( G+\varepsilon E\right) =C_2C,\,\,\,\varepsilon C_0F=C_2D.
\label{enlace de constantes}
\end{eqnarray}
To see the correspondence between our solution and the BTZ solution~(\ref
{BTZ}) we set $G=1$ 
%because our solution is invariant under the transformation
%$e^{2 \gamma} \longrightarrow G^{2} \, e^{2 \gamma}$ and $d \theta
%\longrightarrow G^{-1} \, d\theta$.
and impose the coordinate condition 
\begin{eqnarray}
Ce^{2\sqrt{-\Lambda }\,r}+D=\rho ^2.  \label{cambio de coordenadas}
\end{eqnarray}
From~(\ref{cambio de coordenadas}) and $dt\longrightarrow \sqrt{-\Lambda }%
B \, dt $ we have 
\begin{eqnarray}
ds^2 =(-\Lambda )(\rho ^2-D)^2e^{-2\tilde \gamma }dt^2-
((-\Lambda )(\rho ^2-D)^2)^{-1}\rho ^2d\rho ^2e^{2\tilde \gamma }\left(
d\theta +f(r)e^{-2\tilde \gamma }dt\right) ^2,
\end{eqnarray}
where 
\begin{eqnarray}
f(r) &=&\frac{\kappa C_0C_2\varepsilon C}{4\pi \sqrt{-\Lambda }}\,ln(\frac{%
\rho ^2-D}C)+E\sqrt{-\Lambda }\rho ^2+
\sqrt{-\Lambda }(FC-ED),\hspace{1.5cm}  \nonumber
\end{eqnarray}
and 
\begin{eqnarray}
e^{2\tilde \gamma }=\frac{\kappa C_0^2}{4\pi (-\Lambda )}\,ln(\frac{\rho ^2-D%
}C)+\rho ^2.  \nonumber
\end{eqnarray}
From~(\ref{enlace de constantes}) we can derive $FC-ED=\varepsilon D$ and
write the solution in the following form: 
\begin{eqnarray}
ds^2=(-\Lambda )(\rho -D/\rho )^2e^{-2\Gamma }dt^2- 
\frac{d\rho ^2}{(-\Lambda )(\rho -D/\rho )^2}-\rho ^2e^{2\Gamma }\left(
d\theta +\tilde f(r)e^{-2\Gamma }dt\right) ^2,  \label{metrfin}
\end{eqnarray}
where 
\begin{eqnarray}  \label{f(r)}
\tilde f(r) &=&\frac{\kappa C_0C_2\varepsilon }{4\pi (-\Lambda )\,\rho ^2}%
\,ln(\sqrt{-\Lambda }(\rho ^2-D))+E\sqrt{-\Lambda }+ 
\varepsilon D\sqrt{-\Lambda }/\rho ^2,\hspace{2.5cm}  \nonumber
\end{eqnarray}
and 
\begin{eqnarray}
e^{2\Gamma }=\frac{\kappa C_0^2}{4\pi (-\Lambda )\rho ^2}\,ln(\sqrt{-\Lambda 
}(\rho ^2-D))+1,  \nonumber
\end{eqnarray}
with the condition 
\begin{eqnarray}  \label{ultimo enlace}
C_2=C_0\sqrt{-\Lambda }(1+\varepsilon E)
\end{eqnarray}
for the constants.

The (2+1) local electromagnetic field is given by $E_r=\varepsilon
C_2e^\beta $ and $B=C_0e^\beta $. Thus the electromagnetic potential has the
form 
\begin{eqnarray}  \label{ultimo potencial}
A=\frac{C_0}{2\sqrt{-\Lambda }}\left( ln(\rho ^2-D)\right) \left[ \sqrt{%
-\Lambda }(\varepsilon +E)dt+d\theta \right] .
\end{eqnarray}

We can switch off the electromagnetic field setting $C_0=0$ (this means that 
$C_2=0$ too). Then we see that our spacetime takes the form of the spinning
BTZ solution. The comparison with~(\ref{BTZ}) leads us to $J=2D\sqrt{%
-\Lambda }$. From this we get 
\begin{eqnarray}
(-\Lambda )(\rho -D/\rho )^2=(\sqrt{-\Lambda }\rho -J/2\rho )^2
\end{eqnarray}
or $M=2D(-\Lambda )$, which implies that $J=M/\sqrt{-\Lambda }$. This means
that we have an Einstein-Maxwell generalization of an extreme black hole~\cite%
{Teitelboim1,Teitelboim2}, which has a horizon at $r^2=D=M/2(-\Lambda )$.
From~(\ref{ultimo enlace}) and~(\ref{ultimo potencial}) we obtain that if $%
E=-\varepsilon$ the electric field is switched off and the magnetic field
remains in the found space-time, which takes the form~(\ref{metrfin}) with $%
\tilde f(r) =\varepsilon \sqrt{-\Lambda }( D- \rho ^2)/\rho^{2}$.

It is trivial to see that as $r\longrightarrow \infty $ the line element~(%
\ref{metrfin}) asymptotically does not approach that of the form of the
spinning BTZ solution (\ref{BTZ}). This is because the logarithmic term in $%
\tilde f(r)$ predominates over the term $D\sqrt{-\Lambda }/\rho ^2$ at
spatial infinity. This agrees with the behavior of the angular momentum in
our solution at spatial infinity. To identify the angular momentum in the
metric of the form~(\ref{metrica}) we can use the quasilocal formalism~\cite
{Brown1,Brown2}. Then the angular momentum $j(r)$ at a radial boundary $r$
is given by~\cite{Chan} 
\begin{eqnarray}
j=(e^{\delta -\gamma })^{^{\prime }}\,e^{-\alpha -\beta +3\gamma }.
\label{momento angular 1}
\end{eqnarray}
Taking into account~(\ref{condicion derivada}) and~(\ref{lambda negativo})
the formula~(\ref{momento angular 1}) takes the form 
\begin{equation}
j=\varepsilon (e^{2\gamma })^{^{\prime }}e^{-\beta }-2\varepsilon \sqrt{%
-\Lambda }e^{2\gamma }.  \label{jota}
\end{equation}
Thus for our solution we obtain 
\begin{eqnarray}  \label{mi momento angular}
j &=&\frac{-\kappa \varepsilon C_0^2}{2\pi \sqrt{-\Lambda }}\,ln\left[ \sqrt{%
-\Lambda }(\rho ^2-D)\right] +\frac{\kappa \varepsilon C_0^2}{2\pi \sqrt{%
-\Lambda }}- 2\varepsilon \sqrt{-\Lambda }D.
\end{eqnarray}
When $r\longrightarrow \infty ,$ the angular momentum diverges. Only when $%
C_0=0$ the angular momentum is just finite at spatial infinity (BTZ
solution). 

Finally, it is interesting to clarify the next question: can one generate
our solution from the static Einstein-Maxwell fields with the help of the
combined local coordinate transformation~\cite{Clement} 
\begin{eqnarray}
t\longrightarrow t-\omega \,\theta ,\,\,\,\theta \longrightarrow \theta
-\omega t\text{ }?.  \label{transformacion}
\end{eqnarray}
More exactly, from the electrostatic BTZ (see (\ref{BTZ}) with $A=J=0$ and $%
N^2=L^2=-M-\Lambda r^2-Q_e^2lnr$), with the help of~(\ref{transformacion}),
one can generate a new spinning solution~\cite{Clement}. This field is an
exact solution of the Einstein-Maxwell equations. In this case the
electromagnetic potential may be written as $A=Q_elnr(dt-\omega d\theta )$.
With the help of the same transformation one can generate from the
magnetostatic solution~\cite{Cataldo,Hirschmann} 
\begin{eqnarray}
ds^2=r^2dt^2-g^{-1}(r)dr^2-g(r)d\theta ^2
\end{eqnarray}
a new spinning solution 
\begin{eqnarray}
ds^2=\frac{r^2}{K^2}(1-\omega ^2)^2g(r)dt^2-g^{-1}(r)dr^2-
K^2\left( d\theta -\frac{\omega (g(r)-r^2)}{K^2}dt\right) ^2,
\label{cat}
\end{eqnarray}
where $K^2=g(r)-\omega ^2r^2$ and $g(r)=-M-\Lambda r^2+Q_m^2lnr$, with $%
Q_m^2=\kappa q_m^2/2\,\pi $. In this case the electromagnetic potential is
given by $A=q_m\,lnr(\omega dt-d\theta )$. Since these solutions are
generated from the electrostatic and magnetostatic fields, we can obtain one
from another (and vice versa) with the help of the duality mapping~\cite
{Cataldo} 
\begin{eqnarray}
dt\longrightarrow idt,d\theta \longrightarrow id\theta ,q_m\longrightarrow
iq_e\,\,and\,\,q_e\longrightarrow iq_m.  \label{mapeo dual}
\end{eqnarray}
This means that one can obtain from~(\ref{cat}) and~(\ref{mapeo
dual}) the
spinning charge solution of Cl\'ement~\cite{Clement}. These fields are not
solutions of the Einstein-Maxwell equations~(\ref{Gtt})--(\ref
{Maxwell
condition 2}) obtained by assuming the self (anti-self dual)
equation~(\ref{dual-antidual}). This is easy to see with the help of
equation~(\ref{lambda negativo}), which is not satisfied in this case.
Obviously, they are solutions of the more general equations~(\ref{t-t1})--(%
\ref{t-theta1}) and~(\ref{cond de Max1})--(\ref{cond de Max2})\footnote{%
This was verified by computing the enumerated equations with the help of
MAPLE.}. 

One of the authors (M.C.) would like to thank J.C. Ceballos for helpful
discussions. The authors also thank P. Minning for carefully reading the
manuscript. This work was supported in part by Direcci\'{o}n de
Promoci\'{o}n y Desarrollo de la Universidad del B\'{\i}o-B\'{\i}o through
Grant No 960205-1, and in part by Direcci\'{o}n de Investigaci\'{o}n de la
Universidad de Concepci\'{o}n throug Grant No 96.011.015-1.2.

\end{document}